\documentclass[conference, a4paper]{IEEEtran}
\usepackage[colorinlistoftodos]{todonotes}
\usepackage[utf8]{vietnam}
\usepackage{array}
\usepackage{lipsum}
\usepackage{subfigure}
\usepackage{listings}
\usepackage{xcolor}

\definecolor{codegreen}{rgb}{0,0.6,0}
\definecolor{codegray}{rgb}{0.5,0.5,0.5}
\definecolor{codepurple}{rgb}{0.58,0,0.82}
\definecolor{backcolour}{rgb}{0.95,0.95,0.92}

\lstdefinestyle{mystyle}{
  backgroundcolor=\color{backcolour},   commentstyle=\color{codegreen},
  keywordstyle=\color{magenta},
  numberstyle=\tiny\color{codegray},
  stringstyle=\color{codepurple},
  basicstyle=\ttfamily\footnotesize,
  breakatwhitespace=false,         
  breaklines=true,                 
  captionpos=b,                    
  keepspaces=true,                 
  numbers=left,                    
  numbersep=5pt,                  
  showspaces=false,                
  showstringspaces=false,
  showtabs=false,                  
  tabsize=2
}

\IEEEoverridecommandlockouts
\usepackage{cite}
\usepackage{amsmath,amssymb,amsfonts}
\usepackage{algorithmic}
\usepackage{graphicx}
\usepackage{textcomp}
\usepackage{xcolor}
\def\BibTeX{{\rm B\kern-.05em{\sc i\kern-.025em b}\kern-.08em
    T\kern-.1667em\lower.7ex\hbox{E}\kern-.125emX}}

\begin{document}

\title{Dự đoán chỉ số cường độ tín hiệu thu RSSI \\với các mô hình học máy}

\author{

\IEEEauthorblockN{
Lê Tùng Giang*\IEEEauthorrefmark{2},
Quách Huy Tùng\IEEEauthorrefmark{2},
Đào Lê Thu Thảo\IEEEauthorrefmark{2},
Trần Mạnh Hoàng\IEEEauthorrefmark{2}
} 

\IEEEauthorblockA{\IEEEauthorrefmark{2} 
\textit{Viện Điện tử - Viễn thông, Trường Đại học Bách khoa Hà Nội}\\
Email: \IEEEauthorrefmark{2} \{giang.lt172520, tung.qh156822\}@sis.hust.edu.vn; \{thao.daolethu, hoang.tranmanh\}@hust.edu.vn}
}
\maketitle

\begin{abstract}
Trong nghiên cứu này, chúng tôi đưa ra một phương pháp dự đoán chỉ số cường độ thu (RSSI) trong một khu vực của trạm phát. Các mô hình truyền sóng suy hao truyền thống thường tốn thời gian cũng như độ phức tạp tính toán phụ thuộc nhiều vào yếu tố riêng có của môi trường. Nghiên cứu này tập trung đưa ra giải pháp dự đoán chất lượng tín hiệu sử dụng giá trị tọa độ tại các điểm trong khu vực. Chúng tôi áp dụng các mô hình học máy như là hồi quy tuyến tính, Support Vector Machine (SVM) hay mô hình cây quyết định, để có thể dự đoán trực tiếp chỉ số cường độ tín hiệu thu RSSI của các điểm trong phạm vi của một trạm phát mà không cần tính toán các tham số phức tạp của mô hình truyền sóng suy hao. Hiệu quả của dự đoán RSSI được đánh giá bởi sai số bình phương trung bình (MSE) và sai số tuyệt đối trung bình (MAE). Công đoạn huấn luyện và kiểm thử các mô hình học máy trong nghiên cứu sử dụng dữ liệu là kết quả đo thực tế cúa nhóm trong quá trình nghiên cứu. 
\end{abstract}
\def\IEEEkeywordsname{Từ khóa}
\begin{IEEEkeywords}
RSSI-prediction, machine-learning, USRP, artificial-intelligence
\end{IEEEkeywords}

\section{Giới thiệu}

Ngày nay, để xây dựng một phương án triển khai lắp đặt hệ thống mạng không dây trong nhà hay trong một khu vực, chúng ta phải tính toán và đánh giá rất nhiều tham số ảnh hưởng tới chất lượng của đường truyền bao gồm trễ truyền, mật độ người dùng, độ nhạy máy thu \cite{Nguyen2020}. Ngoài ra chúng ta cũng phải cân nhắc tới chi phí lắp đặt cân đối với hiệu quả sử dụng của dịch vụ. Và trong hệ thống mạng không dây, chúng ta thường sử dụng chỉ số cường độ tín hiệu thu (RSSI) để biểu diễn chất lượng của tín hiệu \cite{Hoa2019EAI}. Với thông tin RSSI tại nhiều điểm trong khu vực đang xem xét, chúng ta có thể xây dựng một bản đồ đường đồng mức năng lượng (Heatmap) biểu diễn trực quan dự đoán về độ phủ sóng của hệ thống. Việc xác định bản đồ đường đồng mức năng lượng sẽ giúp tối ưu hóa vấn đề xây dựng các trạm thu phát, đảm bảo chất lượng đường truyền đạt yêu cầu đưa ra tại tất cả các điểm trong vùng phủ sóng.

Tuy nhiên, việc đo đạc RSSI tại mọi điểm trong khu vực thông thường sử dụng các mô hình suy hao trong môi trường là không đơn giản \cite{ostlin2010}. Các mô hình này thường phụ thuộc nhiều vào các tham số riêng của môi trường, không có tính tổng quát và chỉ tập trung vào một trường hợp nhất định như ngoại ô, đô thị hay môi trường trong nhà \cite{sarkar2003}. Chưa kể việc áp dụng các hiện tượng vật lý như truy vết các tia bức xạ (ray-tracing) \cite{sarkar2003}, lý thuyết tán xạ \cite{brien2000} hay hệ phương trình Maxwell \cite{gorce2007} là vô cùng phức tạp. Nhưng đối với các trường hợp thực tế, các đo đạc, tính toán tham số của môi trường này lại không đáp ứng chính xác và dẫn tới các sai khác trong kết quả của môi trường khác \cite{ayadi2017}. Điều này là bởi các tham số này phụ thuộc rất nhiều vào yếu vật lý của tố môi trường và để đo đạc, tính toán các tham số đó là vô cùng phức tạp. 

Ngoài ra, đối với các thiết bị di động, việc điều khiển công suất phát sẽ được thực hiện nhiều bước qua lại giữa trạm phát và thiết bị để có thể đảm bảo được chất lượng dịch vụ cần thiết \cite{powercontrol}. Phương pháp này không chỉ làm tăng trễ trong hệ thống đặc biệt ảnh hưởng đối với các thiêt bị di chuyển với vận tốc cao, mà còn làm tiêu tốn năng lượng của các thiết bị di động \cite{karmakar2021}. Các mạng viễn thông hiện hành sử dụng tính toán tập trung tại một thiết bị (ví dụ là thục hiện trên cloud). Và với yêu cầu cao hơn về chất lượng cũng như độ trễ thì các bộ tính toán này sẽ khó có thể đáp ứng. Khi đó, công việc tính toán bị dồn nén lại cho các thiết bị tập trung. Điều này tạo ra một lượng trễ rất lớn để thiết lập kết nối với các thiết bị di động, gây ảnh hưởng tới dịch vụ. Ngoài ra, rất dễ gây ra nghẽn mạng nếu số lượng người dùng cũng như lưu lượng tới hệ thống bùng nổ. Vì vậy, tính toán biên sẽ trở thành một công nghệ trọng tâm đối với các công nghệ viễn thông mới \cite{SAFAVAT2020189, elbamby2019}. Đã có rất nhiều công trình nghiên cứu đưa ra dựa trên công nghệ này và đã chứng minh được hiệu quả của nó trong các hệ thóng viễn thông \cite{zhou2020}, trong mạng kết nối vạn vật IoT \cite{raj2021}, và đặc biệt là trong hệ thống mạng 5G \cite{siriwardhana2021} - hệ thống coi trọng giảm thiểu trễ và tính ổn định của hệ thống. Công cụ của chúng tôi cho các trạm phát khả năng tự tính toán công suất phát với độ phức tạp thấp và tốc độ cao để đạt được chất lượng dịch vụ yêu cầu mà không cần thông qua các thiết bị tính toán tập trung. Qua đó giảm trễ cho hệ thống và tăng tính ổn định cho các thiết bị tầng cao hơn. Đồng thời, sử dụng công cụ này sẽ tối ưu hiệu suất thu phát cho các thiết bị ở mọi nơi, đặc biệt là tại rìa các cell.

Như vậy, công cụ này được mong đợi sẽ giúp những người thiết kế mạng có thể dự đoán trước về kế hoạch lắp đặt của mình, tính toán và xem xét được các giới hạn của việc lắp đặt; đồng thời công cụ có thể giúp cho hệ thống giảm độ trễ trong việc xử lý và tính toán công suất phát để đáp ứng chất lượng dịch vụ với độ chính xác cao, tiết kiệm nhân lực và thời gian.


\section{Mô hình hệ thống đo lường và thu thập dữ liệu}
\label{Sec:MoHinhHeThong}


Trong nghiên cứu này, nhóm tác giả huấn luyện các mô hình học máy bằng dữ liệu đo đạc thực tế. Đồng thời, dữ liệu này cũng được dùng để kiểm thử tính chính xác cũng như tốc độ của mô hình sau khi huấn luyện. Huấn luyện các mô hình học máy yêu cầu phải có một tập dữ liệu chính xác và đủ lớn. Vì vậy, việc xây dựng một hệ thống đo đạc và thu thập dữ liệu cho huấn luyện và kiểm thử các mô hình học máy là vô cùng quan trọng.

\subsection{RSSI trong truyền dẫn vô tuyến}


 
Chỉ số cường độ tín hiệu thu (RSSI) là một thuộc tính thể hiện mức công suất của tín hiệu vô tuyến thu được bởi thiết bị tại một khoảng cách so với thiết bị phát. Theo lý thuyết, RSSI là một hàm ảnh hưởng chủ yếu bởi khoảng cách giữa thiết bị thu và phát và các điều kiện khác của môi trường.

RSSI được định nghĩa trong chuẩn IEEE 802.11: “cường độ tín hiệu tương đối nhận được trong môi trường không dây, trong các đơn vị tùy ý”. Trong các môi trường không dây không đồng nhất, khi thiết bị di động cảm nhận được nhiều hơn một mạng không dây cùng một lúc, việc lựa chọn mạng có QoS tốt nhất đóng vai trò quan trọng và tín hiệu nào có RSSI cao đồng nghĩa với chất lượng tín hiệu càng tốt.

Nguyên lý áp dụng của RSSI gồm 3 bước. Đầu tiên máy thu sau khi thu được tín hiệu sẽ đo cường độ tín hiệu và gửi thông tin phản hồi về máy phát. Máy phát nhận thông tin phản hồi và tính toán được trạng thái tín hiệu, khoảng cách thu – phát. Máy phát xử lí tín hiệu phản hồi và so sánh với các ngưỡng công suất tương đường với chất lượng cường độ tín hiệu và thay đổi công suất phát hợp lý.

Dựa vào giá trị của RSSI đọc được từ wifi card, các nhà sản xuất xác định mức chất lượng tín hiệu WLAN và thể hiện qua wifi bar như trong bảng \ref{tab:RSSIstrength}. Khi chỉ số RSSI càng cao thì thể hiện cường độ tín hiệu thu được càng tốt và ngược lại.

\begin{table}[]
    \centering
    \caption{Chất lượng tín hiệu theo chỉ số RSSI}
    \begin{tabular}{|m{0.15\columnwidth}|m{0.15\columnwidth}|m{0.55\columnwidth}|}
        \hline
         \textbf{RSSI} &\textbf{Đánh giá} &\textbf{Mô tả}  \\
         \hline
         -30 dBm & Tuyệt vời & Cường độ tín hiệu tối đa có thể đạt được. Thiết bị ở rất gần điểm truy cập. Thường khó đạt được trong điều kiện thực tế.    \\
         \hline
         -67 dBm  & Rất tốt & Chất lượng tín hiệu yêu cầu tối thiểu cho các ứng dụng yêu cầu độ tin cậy và thời gian truyền kịp thời của gói tin.  \\
         \hline
         -70 dBm & Tốt & Cường độ tín hiệu tối thiểu để đảm bảo độ tin cậy cho việc truyền gói tin. \\
         \hline
         -80 dBm & Yếu & Giá trị cường độ tín hiệu tối thiểu cho các kết nối cơ bản. Gói tin có thể bị mất.\\
         \hline
         -90 dBm & Quá yếu & Mức cường độ tín hiệu thu quá thấp, không thể đảm bảo khả năng kết nối của người dùng.\\
         \hline
         
    \end{tabular}
    \label{tab:RSSIstrength}
\end{table}




\subsection{Xây dựng hệ thống đo đạc}

\begin{figure}[h]
    \centering
    \includegraphics[width = \columnwidth]{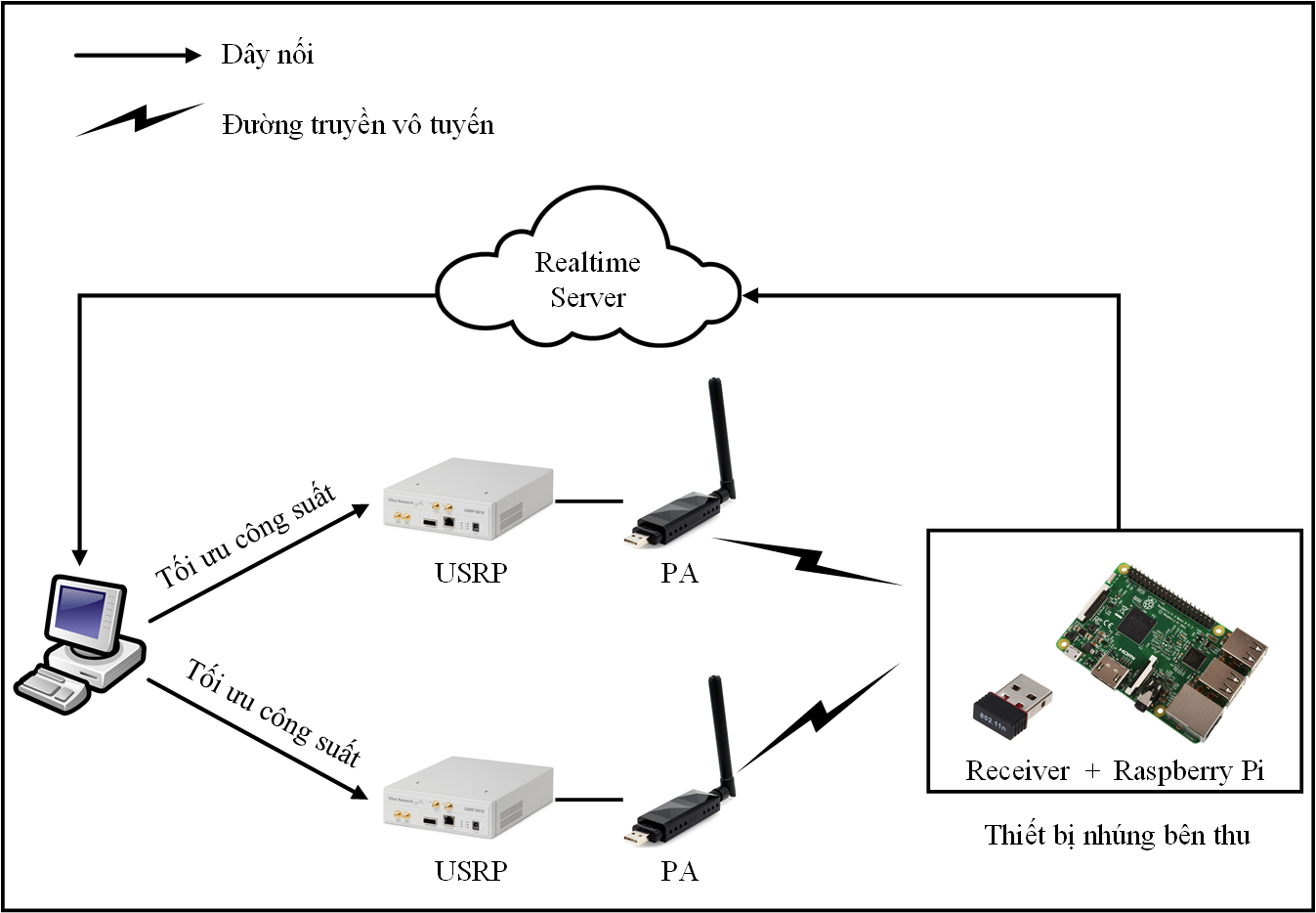}
    \caption{Sơ đồ khối của hệ thống}
    \label{fig:blockSystem}
\end{figure}

Mô hình hệ thống đo đạc sẽ gồm 2 phần chính là phần thu và phần phát như thể hiện trong hình \ref{fig:blockSystem}. Tại bên phát, USRP được lập trình để phát đi liên tục các bản tin với chu kỳ 100ms theo chuẩn IEEE 802.11g (WiFi). Bên thu sử dụng Raspberry Pi 3 với module wifi và di chuyển ngẫu nhiên trong vùng phủ sóng của USRP. Thông qua việc đọc các bản tin được gủi, Raspberry sẽ cập nhật liên tục thông tin về RSSI cũng như thông tin về vị trí đạt được qua module GPS. Đồng thời, tại bên phát, máy tính cũng theo dõi giá trị RSSI này và điều chỉnh để bên thu nhận được tín hiệu tối ưu thỏa mãn chất lượng dịch vụ. Server thời gian thực được sử dụng trong kịch bản là Firebase Realtime của Google.

Máy tính khi nhận được các giá trị RSSI sẽ so sánh với giá trị ngưỡng là -60dBm. Bất cứ khi nào RSSI nhỏ hơn giá trị ngưỡng này thì máy tính sẽ tăng công suất phát của USRP và nếu RSSI lớn hơn giá trị ngưỡng chất lượng là -40dBm (tức tín hiệu quá tốt như trình bày ở bảng \ref{tab:RSSIstrength}) thì máy tính điều khiển cho USRP giảm công suất phát để tránh lãng phí tài nguyên.

\begin{figure}[h]
    \centering
    \includegraphics[width = \columnwidth]{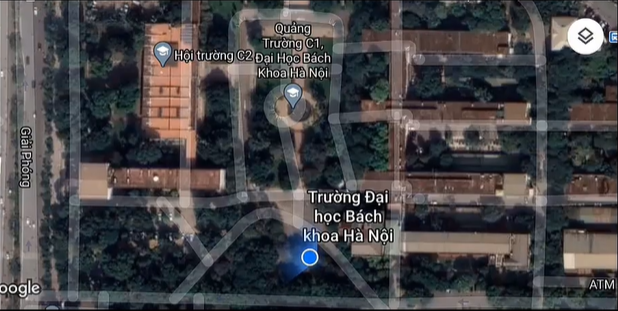}
    \caption{Phạm vi khu vực thực hiện đo đạc}
    \label{fig:DHBKHN}
\end{figure}

\begin{figure*}[t]
    \centering
    \subfigure[Hệ thống lắp đặt bên phát tín hiệu]
    {
        \includegraphics[width=0.45\textwidth]{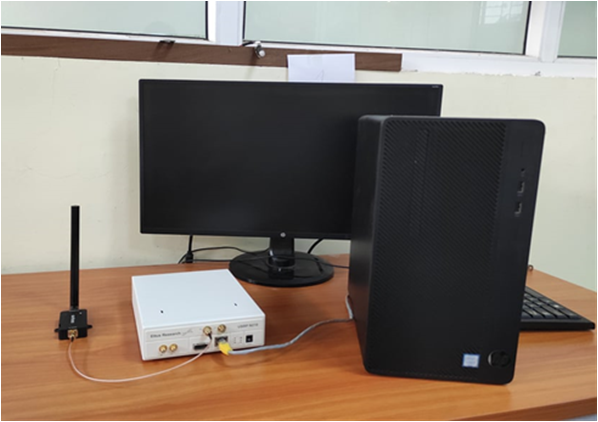}
         \label{fig:Tx}
    } 
    \subfigure[Hệ thống lắp đặt bên thu tín hiệu]
    {
        \includegraphics[width=0.45\textwidth]{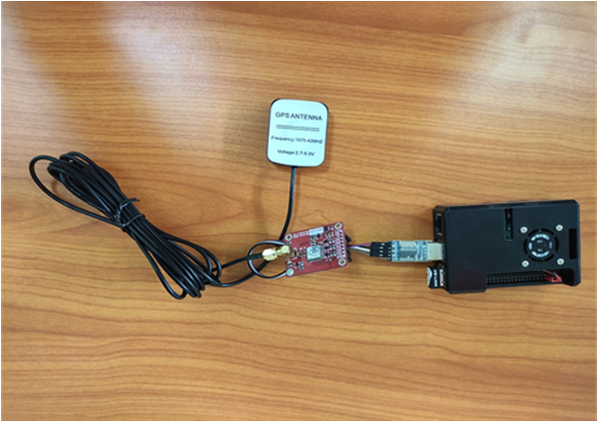}
        \label{fig:Rx}
    }

    \caption{Hệ thống đo đạc thực tế}
    \label{fig:RealMeas}
\end{figure*}

Hình \ref{fig:RealMeas} là hình ảnh hệ thống đo đạc thực tế được nhóm sử dụng trong nghiên cứu này. Hệ thống gồm 2 phần thu và phát như được mô tả ở trên. Phần phát gồm máy tính kết nối với bộ USRP có gắn khuếch đại để điều khiến và phát tín hiệu trong khu vực. Phần thu bao gồm một máy tính nhúng (Raspberry Pi 3) với WiFi có chế độ monitor mode có thể đọc được RSSI của tín hiệu nhận được kết hợp với module GPS để thu lại vị trí. Một người dùng di chuyển chậm xung quanh điểm phát nói trên một cách ngẫu nhiên nhưng phải bao phủ phạm vi xung quanh điểm phát đó. Hình \ref{fig:DHBKHN} là khu vực nhóm thực hiện đo đạc là khuôn viên C9 trường Đại học Bách khoa Hà Nội. 

\subsection{Kết quả đo đạc và xử lý dữ liệu thu được}
Ở trong dự án này, nhóm tác giả đưa dữ liệu thu được qua mô hình xây dựng ở phần \ref{Sec:MoHinhHeThong} vào trong các mô hình học máy khai phá dữ liệu được nghiên cứu ở phần \ref{Sec:DataMining}. Từ đó huấn luyện ra được các mô hình nhận đầu vào là các giá trị tọa độ trong vùng phủ sóng và đầu ra là chỉ số chất lượng RSSI tại tọa độ vừa nhập.
\begin{table}[h]
    \centering
    \caption{Tập dữ liệu thu được sau quá trình đo đạc}
    \begin{tabular}{|m{0.05 \columnwidth}|m{0.12 \columnwidth}|m{0.13 \columnwidth}|m{0.24 \columnwidth}|m{0.2 \columnwidth}|}
        \hline
        \textbf{Mẫu}	&\textbf{Vĩ độ}	    &\textbf{Kinh độ}    &\textbf{Khoảng cách (m)}    &\textbf{RSSI (dB)} \\
        \hline
        1	&21005246	&105842200	&88.17319887	&-97.39850833 \\
        \hline
        2	&21005813	&105841779	&130.1120559	&-113.8386084 \\
        \hline
        3	&21005944   &105841869	&125.6891235	&-111.2070225 \\
        \hline
        4	&21005292	&105843075	&27.45889783	&-57.12242927 \\
        \hline
        5	&21005124	&105841893	&122.7996779	&-114.4533891 \\
        \hline
        ... &...        &...        &...            &... \\
        \hline
    \end{tabular}
    \label{tab:data}
\end{table}

\begin{figure}[h]
    \centering
    \includegraphics[width = \columnwidth]{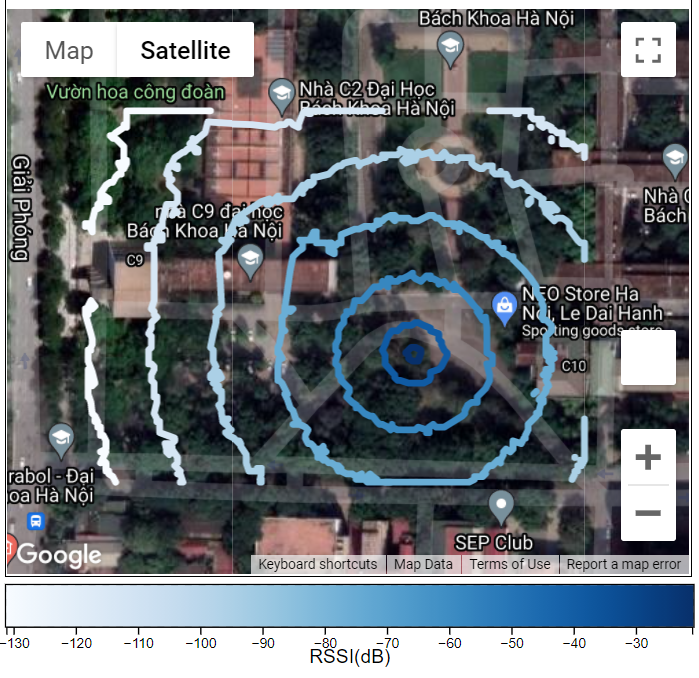}
    \caption{Bản đồ đường đồng mức chất lượng tín hiệu vô tuyến theo RSSI}
    \label{fig:contour}
\end{figure}

Với mô hình hệ thống vừa được trình bày, nhóm đã thu được kết quả bao gồm kinh độ, vĩ độ (được nhân lên $10^6$), khoảng cách và giá trị RSSI tại vị trí đo sẽ được lưu lại với chu kỳ 5 giây 1 lần với số lượng 10.000 mẫu trong file .CSV như được biểu diễn ở bảng \ref{tab:data}. Kết quả đo đạc tất cả 10.000 mẫu được biểu diễn bởi biểu đồ đường đồng mức chất lượng tín hiệu như ở hình \ref{fig:contour}.

Có thể thấy tại các điểm càng gần vị trí phát thì chất lượng tín hiệu càng mạnh, được biểu hiện bằng chỉ số RSSI màu xanh đậm $(-30 dB)$, và khi di chuyển càng ra xa thì chất lượng chỉ số RSSI càng giảm dần. Điều này là dễ hiểu bởi càng đi xa thì suy hao do không gian truyền cũng như trong môi trường càng lớn, làm giảm chất lượng tín hiệu.



Dễ thấy giá trị kinh độ và vĩ độ sai khác rất ít và nếu sử dụng dữ liệu này để huấn luyện thì kết quả sẽ không khả quan. Vì vậy nhóm tác giả sẽ phải thực hiện xử lý phần dữ liệu này.  Vấn đề giá trị kinh độ và vĩ độ sai khác rất ít ở trên là do phạm vi nhóm thực hiện đo đạc là trong khuôn viên trường Đại học Bách khoa Hà Nội. Ở đây, nhóm tác giả chuẩn hóa bằng cách sử dụng công thức sau:
\begin{align}
    l^* = (l \: \text{mod} \: 10.000) / 10.000
\end{align}
trong đó $l^*$ là kinh độ và vĩ độ sau khi chuẩn hóa, $l$ là giá trị kinh độ và vĩ độ sau khi đo đạc, và $mod$ là phép chia lấy phần dư.

Sau quá trình chuẩn hóa, chúng tôi sẽ chia tập dữ liệu ban đầu thành 2 tập con cho mục đích huấn luyện với kích thước là 4.000 mẫu và kiểm thử các mô hình với kích thước là 6.000 mẫu.

\section{Mô hình học máy dự đoán RSSI}
\label{Sec:DataMining}

\subsection{Khai phá dữ liệu trong viễn thông}

Khai phá dữ liệu (Data mining) là quá trình phân loại và sắp xếp các tập dữ liệu lớn, qua đó, xác định các mẫu và thiết lập ra các mối quan hệ nhằm giải quyết một số vấn đề bằng cách phân tích các mẫu dữ liệu. Và trong viễn thông, khối lượng dữ liệu cần xử lý là vô cùng lớn, vì vậy khai phá dữ liệu trở nên vô cùng hữu ích. Trong đề tài này, nhóm sẽ phân tích tập dữ liệu đo được ở \ref{Sec:MoHinhHeThong}, để từ đó suy ra được mối quan hệ giữa tọa độ các điểm trong vùng phủ của USRP với RSSI thu được.

Quá trình khai phá dữ liệu là một quá trình phức tạp bao gồm nhiều bước không chỉ về tính toán mà còn phải chuyển đổi, làm sạch, tích hợp dữ liệu các mẫu và trích xuất ra để đưa vào phân tích. Trong nghiên cứu này, nhóm tác giả sau khi đo đạc thu thập dữ liệu sẽ cần thực hiện một số công đoạn quan trọng. Đầu tiên là tích hợp và làm sạch dữ liệu, dữ liệu sau khi thu được sẽ phải chuẩn hóa và lọc sao cho không có các nhiễu hay bất thường và tích họp lại thành một tập dữ liệu hoàn chỉnh. Thứ hai là tách dữ liệu thành hai tập con là tập huấn luyện và tập kiểm thử cho hai mục đích tương ứng. Tiếp theo, dữ liệu sẽ được sử dụng để trích xuất ra dữ liệu hữu ích mới, ở đây chính là quá trình học máy sẽ được trình bày ở phần sau. Và cuối cùng, dữ liệu sẽ được trình bày, thể hiện một cách trực quan dưới các dạng bảng biểu, hình vẽ hay đồ thị - cụ thể trong nghiên cứu này là sủ dụng bản đồ đường đồng mức (heatmap).

\subsection{Tiêu chí đánh giá kết quả các mô hình học máy}
Trong dự án này, nhóm tác giả sử dụng 2 tham số là MAE và MSE để đánh giá hiệu quả của các mô hình Machine Learning.

\subsubsection{MAE - Mean Absolute Error} (tiếng Việt: "sai số tuyệt đối trung bình") đo độ lớn trung bình của các sai số trong một tập hợp kết quả dự đoán mà không tính đến hướng của chúng. Đây là giá trị trung bình trên mẫu thử nghiệm về sự khác biệt tuyệt đối giữa dự đoán và quan sát thực tế, trong đó tất cả các khác biệt riêng lẻ có trọng số bằng nhau. Phương trình tính toán của MAE là:
\begin{align}
    {\rm MAE} = \frac{1}{n} \sum_{i=1}^{n}  {|Y_i - \hat{Y_i}|}
\end{align}\\
Trong đó $Y_i$ là kết quả thu được của dữ liệu và $\hat{Y_i}$ là giá trị dự đoán.

\subsubsection{MSE - Mean Square Error} (tiếng Việt dịch là: "sai số toàn phương trung bình") của một phép ước lượng là trung bình của bình phương các sai số, tức là sự khác biệt giữa các ước lượng và những gì được đánh giá. MSE là một hàm rủi ro, tương ứng với giá trị kỳ vọng của sự mất mát sai số bình phương hoặc mất mát bậc hai. Sự khác biệt xảy ra do ngẫu nhiên, hoặc vì các ước lượng không tính đến thông tin có thể cho ra một ước tính chính xác hơn. Phương trình tính toán MSE là:
\begin{align}
    {\rm MSE} = \frac{1}{n}  \sum_{i=1}^{n}  {(Y_i - \hat{Y_i})}^2
\end{align}\\
Trong đó $Y_i$ là kết quả thu được của dữ liệu và $\hat{Y_i}$ là giá trị dự đoán.

Việc bình phương giá trị trung bình của sai số sẽ giúp sai số có thể đạo hàm và từ đó có thể áp dụng một số phương pháp tìm điểm tối ưu.

\subsection{Các mô hình học máy sử dụng trong đề tài}
Để phục vụ cho quá trình khai phá dữ liệu trong dề tài này, nhóm tác giả đã thực hiện sử dụng một số các mô hình học máy. Học máy (Machine Learning) là một lĩnh vực nhỏ của trí tuệ nhân tạo (AI), nó khiến cho các máy tính có khả năng tự học hỏi đựa trên các tập dữ liệu đưa vào mà không cần con người lập trình cụ thể. Trong phần này, chúng tôi sẽ đưa ra một số các mô hình học máy được triển khai trong đề tài này như hồi quy tuyến tính, cây quyết định.

\subsubsection{Linear Regression}
\newcommand{\argmin}{\operatornamewithlimits{argmin}}
Hồi quy tuyến tính Linear Regression là một mô hình kinh điển trong học máy, trong đó ta có bộ dữ liệu gồm n ví dụ $(x_1, x_2, x_3, ..., x_n)$ và giá trị cần dự đoán $(y_1, y_2, y_3, ..., y_n)$. Mục tiêu của mô hình là đưa ra một hàm dự đoán tuyến tính $\hat{y} = f(x)$ mà giá trị dự đoán $\hat{y}$ gần với y. Hàm đó có dạng:
$$
  f(x) = \theta'x + \theta_0
$$
Ta muốn hàm $f$ càng khớp với bộ dữ liệu càng tốt. Nghĩa là sai số dự đoán $\hat{y_i}$ và $y_i$ trên điểm dữ liệu $x_i$ càng nhỏ càng tốt. Chúng ta xây dựng hàm mất mát $J(\theta)$ là trung bình của tổng bình phương giá trị sai lệch trên từng điểm dữ liệu trên tập huấn luyện \cite{book:LR}.
\begin{align}
  J(\theta) &= \frac{1}{2}\sum(y_i - \hat{y_i})^2
\end{align}
Để mô hình dự đoán càng gần với các điểm dữ liệu trên tập huấn luyện, ta cần tối thiểu hóa hàm lỗi trên, dẫn đến bài toán tối ưu sau:
\begin{align}
	\min{J(\theta)} = \frac{1}{2}\sum(y_i - \hat{y_i})^2
\end{align}
Khi đó nghiệm của bài toán là:
\begin{align}
	\hat{\theta} = \argmin_\theta J(\theta)
\end{align}

\subsubsection{Support Vector Regression}
Support Vector Machine là một công cụ học máy phổ biến cho các bài toàn phân loại và hồi quy, được phát minh bởi Vladimir Vapnik và cộng sự vào năm 1992. SVM regression được coi là một kỹ thuật nonparametric vì nó dựa vào hàm kernel \cite{book:SVM}. Giải thuật SVR còn được gọi là episilon-insensitive SVM($\epsilon$-SVM), hay L1-loss. Trong giải thuật $\epsilon$-SVM, tập các điểm dữ liệu huấn luyện bao gồm các ví dụ $x_n$ và giá trị đích $y_n$. Mục tiêu là tìm một hàm f sao cho $f(x_n)$ sai khác với $y_n$ một khoảng nhỏ hơn $\epsilon$.

Hàm mục tiêu của mô hình SVM sẽ có dạng:
\begin{align}
    \mathcal{L}(\mathbf{w}) = \frac{1}{2}{||\mathbf{w}||_2^2} + C \sum_{n=1}^N \xi_n
\end{align}
trong đó, $\mathbf{w}$ là ma trận trọng số, các tham số của mô hình học máy để tính toán và phân loại đầu ra theo đầu vào, $C$ là tham số regularization của mô hình SVM, và $\xi$ là sai khác giữa đầu ra thực tế và đầu ra dự đoán của mô hình.

Nếu $C$ nhỏ, thuật toán sẽ điều chỉnh sao cho kết quả dự đoán có margin lớn nhất có thể, tuy nhiên điều này sẽ dẫn tới giá trị $\sum_{n=1}^N \xi_n$ lớn, tức các giá trị dự đoán không sát với các giá trị thực tế. Nếu C lớn, mô hình sẽ cố gắng tối thiểu hóa sai khác giữa đầu ra dự đoán và đầu ra thực tế. Điều này có thể gây ra hiện tượng overfitting, khi đó, mô hình có thể sẽ không đem lại kết quả cao đối với tập dữ liệu khác trong thực tế.

Nhóm tác giả thực hiện tối ưu mô hình với tham số chính là $C$ ảnh hưởng tới việc mô hình sẽ huấn luyện khớp theo tập dữ liệu để đạt hàm mất mát là nhỏ nhất, hay cố gắng tạo ra khoảng bảo vệ nhỏ nhất cho tập dữ liệu như đã được trình bày ở trên.

\subsubsection{Decision Tree Regression}
Cây quyết định được sử dụng để xây dựng cả mô hình hồi quy hoặc phân loại với mô hình dạng cây. Nó chia nhỏ một tập dữ liệu thành các tập nhỏ hơn, trong khi đồng thời từng bước xây dựng cây quyết định tương ứng. Kết quả là là một mô hình cây với các nút quyết định và nút lá. Một nút quyết định có tối thiểu 2 nhánh, mỗi nhánh đại diện cho giá trị của thuộc tính được kiểm tra tại nút đó. Nút lá đại diện cho giá trị mục tiêu, có thể là cả dạng nhãn ( đối với các bài toán phân loại), hoặc giá trị số (đối với các bài toán hồi quy) \cite{book:DT}.

Giải thuật chính để xây dựng cây quyết định là ID3, C4.5, và CART. Ở đây, ta khảo sát giải thuật ID3, là một giải thuật được phát triển bởi J.R.Quinlan, xây dựng cây theo mô hình top-down, tìm kiếm tham lam trên không gian các nhánh có thể phân chia và không quay lui. Trong khi giải thuật ID3 cho bài toán phân loại sử dụng độ đo Information Gain, thì với bài toán hồi quy ID3 sử dụng độ đo sự suy giảm độ lệch chuẩn (standard deviation reduction) để xây dựng cây.

 Các tham số mà nhóm tác giả thực hiện tối ưu trong công trình này là:
\begin{itemize}
    \item Độ sâu tối đa của cây: Nhóm triển khai các mô hình với giá trị độ sâu là [20, 50, 80, 100, 300, 500, 1000]
    \item Số mẫu tối thiểu để phân nhánh: triển khai tối ưu với các giá trị [2, 5, 10]
    \item Số mẫu tối thiểu ở mỗi nút lá: triển khai tối ưu với các giá trị [2, 5, 10].
\end{itemize}

Ở đây, các tham số tối ưu của cây quyết định chủ yếu là các điều kiện dừng của mô hình. Điều kiện dừng trong mô hình này có tác dụng hạn chế việc overfitting trong huấn luyện, giúp dừng lại thuật toán trước khi hàm mất mát đạt giá trị quá nhỏ, khiến cho mô hình áp dụng quá khớp với tập huấn luyện mà không áp dụng được cho các tập dữ liệu khác trong thực tế.


\subsubsection{Random Forest Regression}
Giải thuật Random Forest tăng độ chính xác của mô hình bằng cách sử dụng nhiều cây quyết định, bằng kĩ thuật bagging. Khi ra quyết định, mô hình Random Forest sẽ lấy trung bình giá trị dự đoán trên các cây trong rừng. Ý tưởng chính của giải thuật như sau: Ở mỗi lần phân chia cây, một tập ngẫu nhiên m thuộc tính được lấy ra và chỉ m thuộc tính này tham gia vào việc phân chia cây. 

Đối với mỗi cây phát triển dựa vào một mẫu bootstrap, tỷ lệ lỗi của các phần tử không thuộc vào bootstrap được kiểm soát, gọi là tỷ lệ lỗi out-of-bag (OOB).Dữ liệu out-of-bag được sử dụng để ước lượng lỗi tạo ra từ việc kết hợp các kết quả từ các cây, tổng hợp trong giải thuật Random Forest, cũng như để ước lượng độ quan trọng của thuộc tính.

Các tham số tối ưu của mô hình này giống với các tham số trong mô hình cây quyết định mà nhóm tác giả đã thực hiện ở trên, tuy nhiên bổ sung thêm một tham số là Số lượng cây quyết định trong rừng ngẫu nhiễn do bản chất mô hình rừng ngẫu nhiễn là tập hợp của một số lượng cây quyết định. Ở đây, nhóm tác giả thực hiện tối ưu với giá trị số lượng cây quyết định là [10, 20, 50, 100, 300, 500].

\subsubsection{Gradient Boosting Tree}
Tương tự như Random Forest, Gradient Boosting Tree cũng là một mô hình được xây dựng từ nhiều cây quyết định. Tuy nhiên như đã trình bày ở trên thì Rừng ngẫu nhiên thuộc một nhóm gọi là các thuật toán Bagging. Các model trong Bagging đều là học một cách riêng rẽ, không liên quan hay ảnh hưởng gì đến nhau, điều này trong một số trường hợp có thể dẫn đến kết quả tệ khi các model có thể học cùng ra 1 kết quả. Chúng ta không thể kiểm soát được hướng phát triển của các model con thêm vào bagging.

Thay vào đó thì Gradient Boosting hay điển hình với Gradient Boosting Tree thuộc nhóm Boosting, tức là các mô hình con sẽ lấy kết quả của nhau và phát triển thêm. Ý tưởng của phương pháp này là tìm các cực tiểu cục bộ của bài toán rồi từ đó tìm được cực tiểu toàn cục - điều này là vô cùng khó đối với một mô hình riêng rẽ (decision tree). Tuy nhiên, do Boosting là một quá trình tuần tự, không thể xử lí song song, do đó, thời gian train mô hình có thể tương đối lâu.

Tương tự như rừng ngẫu nhiên là tập hợp của nhiều cây quyết định nên các tham số trong mô hình này tương đồng với các tham số của mô hình rừng ngẫu nhiên đã được trình bày ở trước.

\section{Kết quả và thảo luận}
\label{Sec:KetQua}

\begin{table}[t]
    \centering
    \caption{So sánh kết quả triển khai các mô hình học máy}
    \begin{tabular}{|m{0.35\columnwidth}|m{0.25\columnwidth}|m{0.25\columnwidth}|}
        \hline
        \textbf{Mô hình}            &\textbf{MAE (dB)}    &\textbf{MSE (dB)}  \\
        \hline
        Linear Regression           &1.8481     &6.2867 \\
        \hline
        Support Vector Regression   &1.0756     &2.8131 \\
        \hline
        Decision Tree               &0.8846     &1.4448 \\
        \hline
        Random Forest Regression    &0.7699     &1.0072 \\
        \hline
        Gradient Boosting Tree      &0.8009     &1.1559 \\
        \hline    
    \end{tabular}
    \label{tab:compare}
\end{table}

Từ kết quả triển khai các mô hình, ta so sánh hiệu quả của các mô hình như trong bảng \ref{tab:compare}. Có thể thấy Random Forest Regression đem lại MSE và MAE nhỏ nhất, tức là mô hình này đem lại kết quả tốt nhất trên tập kiểm tra.

\begin{table}[t]
    \centering
    \caption{So sánh thời gian huấn luyện các mô hình học máy}
    \begin{tabular}{|m{0.45\columnwidth}|m{0.45\columnwidth}|}
        \hline
        \textbf{Mô hình}            &\textbf{Thời gian huấn luyện (phút)}    \\
        \hline
        Linear Regression           &0.1     \\
        \hline
        Support Vector Regression   &3.14     \\
        \hline
        Decision Tree               &0.5    \\
        \hline
        Random Forest Regression    &14.1    \\
        \hline
        Gradient Boosting Tree      &126.81    \\
        \hline    
    \end{tabular}
    \label{tab:time_compare}
\end{table}

Bảng \ref{tab:time_compare} biểu diễn thời gian huấn luyện các mô hình nhóm tác giả thực hiện trong công trình này. Kết hợp với kết quả từ bảng \ref{tab:compare}, ta thấy rằng mô hình rừng ngẫu nhiên vừa đem lại kết quả tốt nhất trên tập huấn luyện, vừa có thời gian huấn luyện thấp. Ngoài ra, nhóm cũng đã kiểm tra thời gian dự đoán của mô hình học máy vào khoảng $10 \: ms$ đối với một mẫu đầu vào gồm kinh độ, vĩ độ và khoảng cách của điểm thu.

\begin{figure}[h]
    \centering
    \includegraphics[width =  \columnwidth]{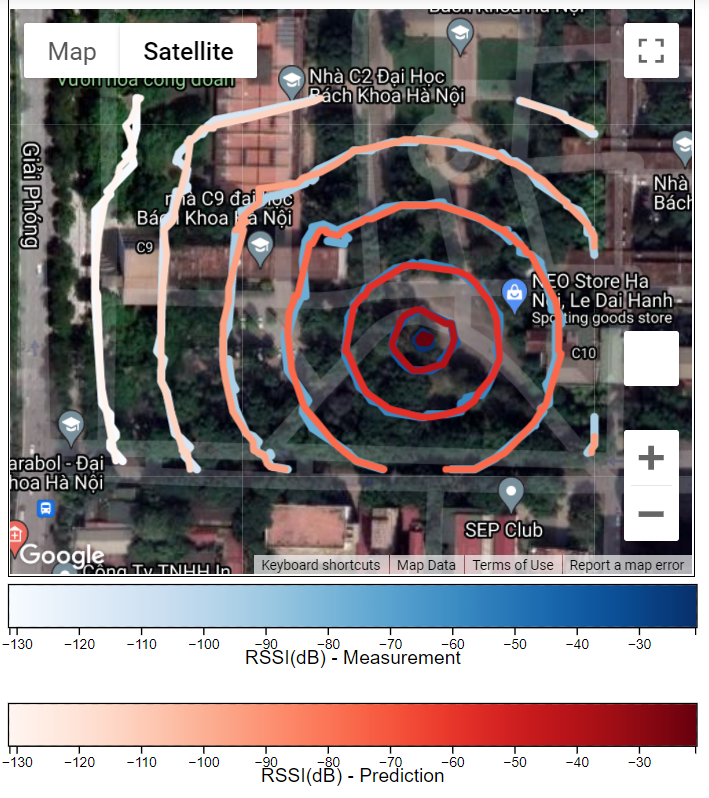}
    \caption{Bản đồ đường đồng mức chất lượng tín hiệu vô tuyến theo RSSI sử dụng kết quả đo đạc và dự đoán}
    \label{fig:result1}
\end{figure}

Hình \ref{fig:result1} là kết quả biểu thị đường đồng mức chất lượng tín hiệu RSSI dựa vào kết quả đo đạc từ phần \ref{Sec:MoHinhHeThong} và kết quả thu được thông qua mô hình rừng ngẫu nhiên dựa trên tập kiểm tra với kích thước là 6.000 mẫu. Các đường đồng mức màu xanh biểu diễn chỉ số RSSI các tham số đo đạc từ phần \ref{Sec:MoHinhHeThong}, còn các đường đồng mức màu đỏ là kết quả thu được từ tập dữ liệu qua mô hình học máy. Có thể thấy kết quả thu được thông qua mô hình học máy huấn luyện từ 4.000 mẫu đạt gần như trùng khớp với kết quả đo đạc về chất lượng tín hiệu tại các vị trí với độ chính xác cao. Từ đó có thể rút ra được nhận xét rằng độ chính xác và ổn định của các mô hình học máy là rất cao và không cần tập dữ liệu huấn luyện với kích thước quá lớn. Kết quả dự đoán từ các mô hình học máy hoàn toàn có thể đáp ứng được các yêu cầu trong thực tế.

\section{Tổng kết}

Chỉ số RSSI là một tham số vô cùng quan trọng đối với các hệ thống truyền thông vô tuyến trong thực tế. Và qua nghiên cứu này, công cụ của chúng tôi sử dụng đầu vào là kinh độ và vĩ độ các điểm trong khu vực, và từ đó đưa qua các mô hình học máy để dự đoán chỉ số RSSI của tín hiệu tại điểm trong khu vực với. Kết quả dự án đem lại cho thấy hiệu quả, đơn giản và độ chính xác cao và trong thời gian ngắn của các mô hình học máy so với kết quả đo đạc thực tế. Với thông tin chỉ số RSSI tại tất cả các điểm trong khu vực, có thể giúp ích nhiều trong nghiên cứu cũng như xây dựng các hệ thống thông tin vô tuyến. Như vậy, công cụ này được mong đợi sẽ giúp thể dự đoán trước về kế hoạch lắp đặt hệ thống, tính toán và xem xét được các giới hạn của việc lắp đặt; đồng thời công cụ có thể giúp cho hệ thống giảm độ trễ trong việc xử lý và tính toán công suất phát để đáp dựng chất lượng dịch vụ với độ chính xác cao, tiết kiệm nhân lực và thời gian.


\bibliographystyle{IEEEtran}

\def\refname{Tài liệu tham khảo}
\bibliography{reference}

\vspace{12pt}

\end{document}